\documentclass[a4paper]{article}
\usepackage[affil-it]{authblk}
\usepackage[dvipdfmx]{graphicx, color}
\usepackage{amssymb}
\usepackage{amsmath}
\usepackage{bm}
\usepackage{cases}
\usepackage{blkarray}
\usepackage{natbib}
\setcitestyle{authoryear, aysep={,}}
\usepackage{lineno}

\usepackage{setspace}

\begin{document}

\title{Towards understanding network topology and robustness of logistics systems}
\date{}
\author[a]{Takahiro Ezaki*}
\affil[a]{Research Center for Advanced Science and Technology, The University of Tokyo, 4-6-1 Komaba, Meguro-ku, Tokyo 153-8904, Japan}
\author[a]{Naoto Imura}
\author[a]{Katsuhiro Nishinari}

\maketitle

\begin{abstract}
	Advanced integration of logistics systems has been promoted for the sake of competitiveness and sustainability. Such efforts will enable more globally optimal and flexible operations by efficiently utilizing transportation capacity. At the same time, interconnection of transport operations increases complexity at a network level, which reduces the predictability of the response of the system to disruptions. However, our understanding of the behavior of such systems is still limited. In particular, the topology of the network, which changes as the systems are integrated, is an important factor that affects the performance of the entire system. Knowledge of such mechanisms would be useful in the design and evaluation of integrated logistics.
	Here, we developed a simple mathematical model that extracts the essence of the problem and performed extensive numerical experiments by Monte Carlo simulations for three scenarios that mimic changes in demand: (i) locally and temporally increased traffic demand, (ii) globally and temporally increased traffic demand, and (iii) permanent change in demand pattern, under various conditions on the type of route-finding algorithm, network structure, and transportation capacity.
	Adaptive route-finding algorithms were more effective in square lattice and random networks, which contained many bypass routes, than in hub-and-spoke networks. Furthermore, the square lattice and random networks were robust to the change in the demand pattern and temporal blockage of delivery paths (e.g., due to high demand).
	We suggest that such preferable properties are only present in networks with redundancy and that the bypass structure is an  important criterion for designing network logistics.
\end{abstract}

*e-mail: tkezaki@g.ecc.u-tokyo.ac.jp 

\clearpage

\section{Introduction}
Logistics systems are becoming increasingly complex and interrelated, from the operational level to the strategic level, due to economic globalization, offshoring of production, increasing product complexity, and fast-changing trends, among others. \citep{Harland2003-ez, Dolgui2020-yf, Choi2001-pl}.
Under these circumstances, companies have been pursuing various types of efficiency improvements for greater competitiveness. One of the main pillars of such measures is the consolidation of logistics systems \citep{Buffa1987-ne, Hall1987-id, Cetinkaya2006-wr}.
Logistics integration is now being practiced not only within companies but also between different companies, i.e., the horizontal collaboration \citep{Chan2004-xx, Cruijssen2007-qq,Naesens2009-mu}.
The horizontal cooperation has been attempted and practiced at various levels. \cite{Pan2019-sm} classified extant horizontal cooperation schemes into six categories, i.e.,  (i) single carrier collaboration \citep{Puettmann2010-jq,Hernandez2011-xq}, (ii) carrier alliance and coalition \citep{Klaas-Wissing2010-yx}, (iii) transport market place \citep{Huang2013-qf}, (iv) flow-controlling entities collaboration \citep{Cruijssen2007-qq}, (v) logistics pooling \citep{Pan2019-sm}, and (vi) Physical Internet (PI) \citep{Ballot2011-hm, Montreuil2011-hh}.
These schemes connect logistics networks of participating parties, generating larger and more complex network structure.
More advanced cooperation generally allows for more global efficiency, but requires integrations of many stakeholders not only at an operational level but also at organizational and managerial levels \citep{Chan2004-xx, Cruijssen2007-qq, Chen2008-yb, Naesens2009-mu,Barthe-Delanoe2014-wz, Da_Silva_Serapiao_Leal2019-xw, Pan2019-zz, Pan2021-ib}.
In addition to these implementation issues, here we raise another issue related to the increased complexity of the system. The transportation dynamics in such interrelated and automated systems and how they respond to changes in demands are difficult to grasp intuitively. While simulating various situations can provide predictive assessments on the performance, sustainability, and robustness of logistics operations, they are often not based on a phenomenological understanding and thus may be sensitive to unanticipated perturbations.We believe that understanding the basic behavior of the entire system in a logistics network is useful for (i) strategic decision making on how to proceed with horizontal cooperation, (ii) scenario design for predictive simulation, and (iii) interpretation of empirical data and simulation results.
However, research based on such an approach is still scarce \citep{Treiblmaier2020-ur}.

Notably, the fields of applied mathematics, statistical physics, and network science have advanced our understanding of the fundamental characteristics of traffic congestion in networks in various contexts \citep{Boccaletti2006-ru, Tadic2007-kk, Chen2011-fy}. Past literature has also underscored the usefulness of such findings in supply chain management \citep{Hearnshaw2013-kf}.
Such theoretical studies have elucidated the emergence mechanisms of congestion in
computer networks \citep{Ohira1998-vk}, road networks \citep{Biham1992-or},
airport networks \citep{Ezaki2014-fi}, and production networks \citep{Ezaki2015-sy}, among others.
Previous studies have found critical determinants of transport performance,
e.g., network topology \citep{Guimera2002-eo,Zhao2005-vk}, capacity distribution \citep{Zhao2005-vk, Wu2008-do}, and routing algorithm \citep{Echenique2004-jw, Zhang2007-vy, Ling2010-dh, Ezaki2015-yu}.
These factors are taken into consideration in this study.

Although such theoretical studies have provided useful implications to various applications,
the conclusions are not directly applicable to the design of logistics networks for the following reasons.
First, to enable mathematical analysis, these studies used models that are too simple (e.g., without  considering delivery costs and optimized route finding) relative to logistics networks.
Second, because the focus of these studies is often on the (very) large-scale behavior of the system, e.g., scale-free properties of networks and phase transitions, the conclusions may not be true or dominant in a realistic system of a relatively small size.
In addition, for this reason, heuristic algorithms with very low computational costs were used as the routing algorithms \citep{Chen2011-fy}.
However, given the distributed computations in logistics networks, high-performance routing should be assumed.

On the other hand, network-level modeling of logistic systems has been performed in the context of the PI, which is a decentralized logistics network, characterized by the use of modularized containers (PI containers) and standardized distribution centers (PI hubs) that connect to each other.
In this context, the typical uses of mathematical models can be divided into three main categories: analytical computation, optimization, and simulation.
For example,  \cite{Ballot2011-hm} considered different types of network topology and evaluated the performance of the PI using an analytical computation based on the continuous approximation method \citep{Langevin1996-ip}.
Optimization studies have shown the efficacy of the PI \citep{Sohrabi2011-nr, Venkatadri2016-nl} and the PI with inventory management
\citep{Pan2015-sa, Ji2019-il}. Compared with these two uses of models, simulation allows for more complex settings,
e.g., optimized route-finding  and dynamic changes in the environment.
Past studies have simulated relatively large-scale systems considering realistic operations \citep{Hakimi2012-xy, Sarraj2014-fk}, a logistics system in a road network \citep{Fazili2017-jy}, disruptions in the network \citep{Yang2017-sh}, and urban logistics operations \citep{Kim2021-gu}. These studies have successfully provided useful insights based on realistic settings, but their approaches are not suitable for drawing out general understanding of the network dynamics.

In contrast to the previous simulation studies of the PI, we use a simple model that abstracts the essence of network logistics to investigate general and fundamental properties of the system, focusing on how the response of the system to various disturbances is affected by the network structure, capacity distribution, and type of (non-)adaptive routing algorithm. We extensively perform numerical experiments based on the model and elucidate the system behavior.
This approach allows us to find and analyze the intrinsic behaviors of logistics network that do not depend on specific details of the system. Of course, we need to be careful that our conclusions are not significantly influenced by factors that we ignored in the modeling process, which will be discussed later.
We examine three types of changes in demand (two temporal changes and one permanent change), and assess the traffic. We find that the network topology significantly affects performance in all cases and show that hub-and-spoke networks are not robust demand changes in general.

The remainder of the article is organized as follows.
First, Sec. \ref{sec:model} defines the examined model and scenarios. Then, Sec. \ref{sec:results} shows the comprehensive simulation results. Finally, we summarize the results and discuss their implications and future prospects in Sec. \ref{sec:conclusion}.

\section{Model}\label{sec:model}
\subsection{Overview of model}
Consider a network of transit centers. Transit centers connected by a link can send freight to each other.
Here, a packet is the minimum unit of delivery, and the link capacity is the maximum number of packets that can be carried at one time on the link. The cost of delivery is also defined for each link.
If demand exceeds capacity, packets must either wait until the demand on that link falls below the capacity or use a bypass route.
We place $N$ packets in the system. When each packet is generated, we randomly choose the origin and destination nodes and compute the route plan by using one of the algorithms explained in Sec. \ref{sec:algo}.
We update the system in discrete time. At each time step, each packet is allowed to jump to the next node if the capacity of the link is not exceeded. When a packet arrives at the destination node, the packet is removed from the system, and a new packet is generated with a randomly chosen origin and destination.

\subsection{Networks}
We consider three types of networks with 25 nodes. In the square lattice network (Fig.  \ref{fig:network_sch}(a)), nodes adjacent to each other on the top, bottom, left, and right are connected together.
On the link, the cost and capacity values are defined in both directions separately; i.e., the cost and capacity values of the link from node 1 to node 2 may be different from those of the opposite link. The cost value of the link from node $i$ to node $j$ ($1\leq i\neq j\leq 25$), denoted by $c_{ij}$, is chosen uniformly at from the interval $[0.5, 1.5]$.
We determine the capacity values via three ways, which are explained in the next subsection.
The lattice has 40 bidirectional links (= 80 directional links).
This type of network is often used to model road networks \citep{Zeng2014-xg}.

In the random network (Fig.  \ref{fig:network_sch}(b)), the 25 nodes are randomly connected by 40 bidirectional links.
To wire the links, we randomly select a pair of non-wired nodes until the total number of links reaches 40.
If the generated network is disconnected (i.e., with isolated parts), then we discard it and regenerate a new random network.
As with the square lattice network, the cost value is randomly drawn from the interval $[0.5,1.5]$ for each direction of each link.
This network is examined to test whether the results are affected by
the regular structures of the two other networks.

The hub-and-spoke network is shown in Fig.  \ref{fig:network_sch}(c).
This network has 25 nodes and 24 bidirectional (48 unidirectional) links.
The cost value is randomly chosen, as with the two other networks.
This type of network is widely used in traditional supply chains for its efficiency \citep{Abdinnour-Helm1999-ke}.

\subsection{Capacity}
\begin{figure*}[t]
	\centering\includegraphics[width=160mm]{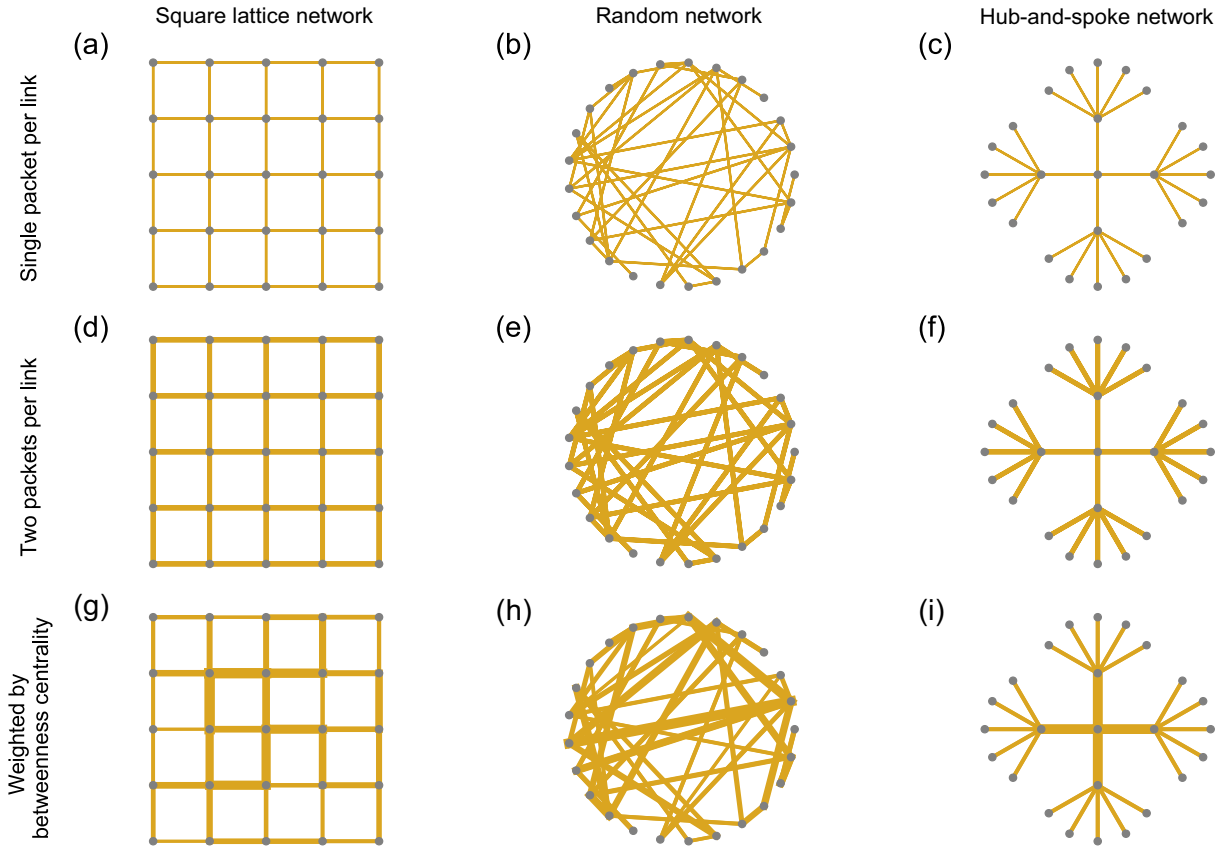}
	\caption{Capacity distributions. (a--c) Single-packet-per-link capacity distribution for each type of network. (d--f) Two-packets-per-link capacity distribution for each type of network. (g--i) Weighted-by-betweenness-centrality capacity distribution for each type of network.
		The thickness of the line representing each link corresponds to the capacity value, $C^{X}_{ij}$ ($1\leq i \neq j \leq 25$, $X=$ single, two, or weighted).
	}
	\label{fig:capacity_sch}
\end{figure*}

The capacity of a link from node $i$ to node $j$ ($1\leq i\neq j\leq 25$), $C_{ij}$, defines the maximum number of packets that
can pass through it at the same time. We use three types of capacity distributions (Fig. \ref{fig:capacity_sch}). The first one, which we call ``\textit{single packet per link},'' defines
the capacity to be $C^{\rm Single}_{ij}=1$ (packet) for all links. The second one, ``\textit{two packets per link},'' doubles this capacity to $C^{\rm Two}_{ij}=2$ (packets) for all links.

The last one, ``\textit{weighted by betweenness centrality},''  weights link capacities according to pre-estimated demand for use.
The demand is estimated by the betweenness centrality \citep{Yoon2006-lp}, which is computed as follows.
First, the shortest paths between all the possible pairs of nodes are determined ($25\times 24=600$ pairs).
Second, for each link from node $i$ to node $j$ ($1\leq i\neq j\leq 25$), the number of shortest paths that use the link is counted and denoted by $n_{ij}$. Note that this definition gives the link (or edge) betweenness centrality, which is a derivative of the commonly used (node) betweenness centrality \citep{Barthelemy2004-bk}.
Finally, the capacity is computed by
\begin{equation}
	C^{\rm Weighted}_{ij} = 1 + \frac{n_{ij}L}{\displaystyle\sum_{i'=1}^{25}\sum_{\substack{j'=1 \\ j'\neq i'}}^{25}{n_{i'j'}}},
\end{equation}
where $L$ is the total number of (unidirectional) links ($L=80$ for the square lattice and random networks and $L=48$ for the hub-and-spoke network).
The sum of the capacity values, $C^{\rm Weighted}_{ij}$, over the links coincides with that of $C^{\rm Two}_{ij}$,
allowing comparisons between the capacity distributions given the total capacity resources.
We use these two distributions, i.e., \textit{the two-packets-per-link} and \textit{weighted-by-betweenness-centrality} capacity distributions, as representatives of a uniform distribution and a distribution weighted by demands, respectively. The \textit{single-packet-per-link} distribution is used as the baseline to be compared with these two distributions to examine the effects of doubling the capacity.

We do not limit the number of packets that nodes can hold.

\subsection{Routing algorithms}\label{sec:algo}
We use three types of routing algorithms: static shortest path (SSP) algorithm, temporary fastest path (TFP) algorithm, and adaptive fastest path (AFP) algorithm.
When a packet enters a system with randomly selected origin and destination nodes, a routing algorithm generates a path plan. Because the availability of a link changes over time due to congestion, a packet may not always be able to travel with the minimum time steps.
A packet using the SSP algorithm always follows the pre-calculated shortest path. If a move is blocked, the packet will wait until the link becomes available.
The TFP algorithm considers the path plans of other existing packets and avoids the blocked link in the future if necessary.
The AFP applies the TFP algorithm every time the situation is changed (e.g., when a new packet is added to the system or the pattern of link availability is updated).
In the SSP and TFP algorithms, the planned path is reserved for the packet and thus not changed on the way to the destination node. In the AFP algorithm, the planned path is updated adaptively according to the environment.

The SSP algorithm follows the shortest path regardless of congestion. If a link in the shortest path is temporally unavailable (e.g., due to excess demand),
the packet will wait until the link becomes available. The shortest path between nodes is computed via Dijkstra's algorithm \citep{Dijkstra1959-is} on the network with cost values.
Note that capacity values were not used to compute the shortest path.

\begin{figure*}[t]
	\centering\includegraphics[width=160mm]{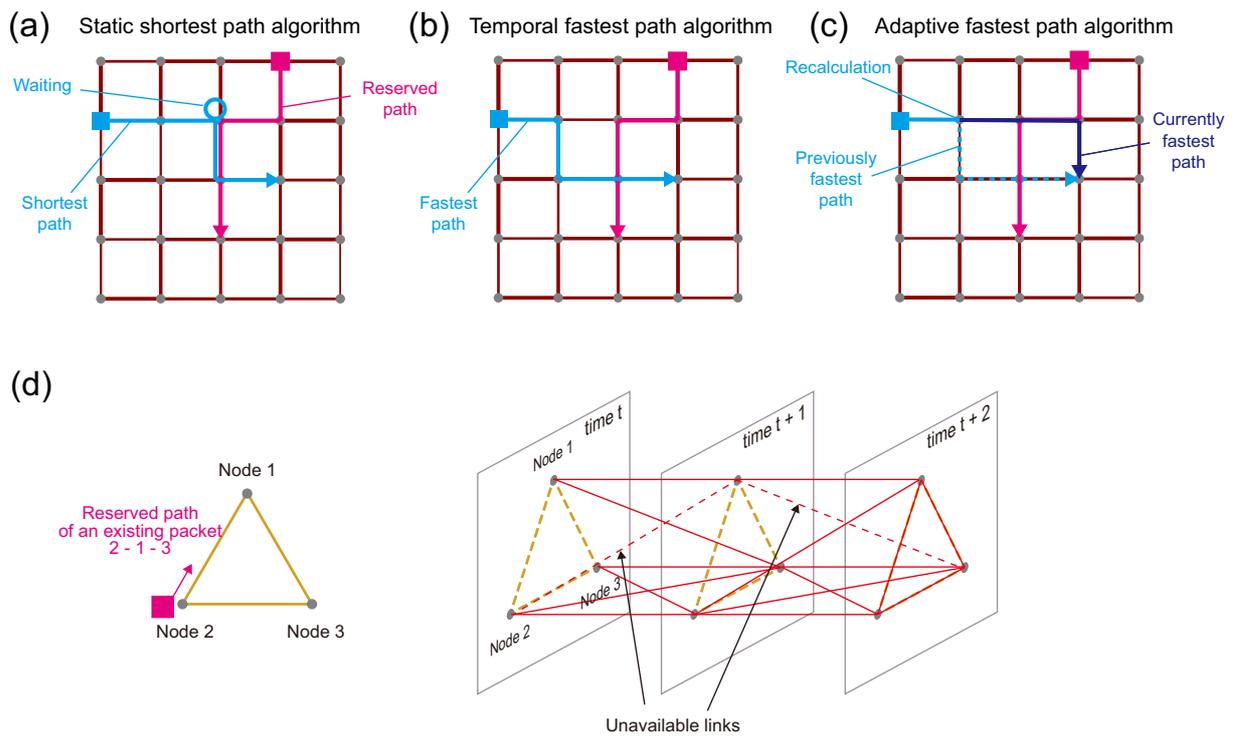}
	\caption{Schematics of routing algorithms. (a) Static shortest path (SSP) algorithm. (b) Temporary fastest path (TFP) algorithm. (c) Adaptive fastest path (AFP) algorithm. (d) Time-expanded-network approach used in the TFP and AFP algorithms.}
	\label{fig:algo_sch}
\end{figure*}

The TFP and AFP algorithms take the number of time steps of travel, including wait time, into consideration.
They find the fastest path, i.e., the path that requires the fewest time steps to travel from the given origin to the destination. (Note that the shortest path is the path with the minimum total cost, but it does not consider the waiting cost, which is explained later.) In our model, because the cost values on the links are not significantly different from each other, the fastest path coincides with the shortest path in many cases when all the links in the network are available. To find the fastest path, we use Dijkstra's algorithm on a time-expanded graph in which a node in a network at different time points is represented by two different nodes (Fig. \ref{fig:algo_sch}). The move from node $i$ to $j$ ($1\leq i\neq j\leq 25$) available at time $t$ is represented by a transition link from node $i$ in time $t$ to node $j$ in time $t + 1$. The cost of using this transition link is $c_{ij}$. If a packet at node $i$ does not move at time $t$ (to wait until a link becomes available), the transition is represented by a transition link from node $i$ in time $t$ to node $i$ in time $t+1$. The cost for using this transition link is defined as $c_{\rm W}$. In this paper, we examine two conditions of the waiting cost: $c_{\rm{W}}=0$ and $c_{\rm{W}}=0.5$. The waiting cost represents a penalty for delayed deliveries, which increases storage cost and reduces user satisfaction.
In this time-expanded network, we compute the minimum cost to reach each of the nodes in each time step from a given node at a given time (nodes that cannot be reached at each time step are also identified in this process). The path that reaches the destination first was is defined as the fastest path.

The technical details of this procedure are as follows. First, we determine the time when the use of the reserved transition links will be completed, $t'$.
We construct a time-expanded graph from the current time $t_0$ to $t'$. If the number of packets that will use a link from $i$ to $j$ at time $t$ is smaller than the link capacity, we place a transition link;
otherwise, we do not. At time $t'$, we also place links between connected nodes, because the standard Dijkstra's algorithm is available for $t\geq t'$. If more than one fastest path (with the same travel time and cost) is found,
the path that approaches the destination in the shortest time is adopted. For example, if routes $1-1-2-3$ and $1-2-2-3$ are found from node 1 to node 3, the latter is adopted as the fastest path.
This happens when the link from node 2 to node 3 is not available until the last time step, in which the choice of waiting at node 1 or node 2 does not affect the cost or time.
Note that although Dijkstra's algorithm \citep{Dijkstra1959-is} is computationally affordable in the systems used in this paper,
it is not applicable when the size of the network becomes large, in which case we need to use more efficient algorithms.

To generate a time-expanded graph, we first identify the links that will be available in the future, considering the already planned paths of existing packets. The TFP algorithm sets the path plan in this way and does not change it.
The AFP algorithm performs these procedures every time a new packet enters the system or a blocked pattern of links is changed.
When such an event occurs, we clear all the planned paths. Then, we re-compute the fastest path for each packet in order of proximity to the destination node (i.e., the number of links in the shortest path from the current position to the destination node). This ordering is based on the premise that (i) a packet far from the destination has a higher possibility of finding a good alternative path than does a packet close to the destination, even when a link along the shortest path is blocked by other packets, and that (ii) such an algorithm generates globally better path plans than those produced without ordering.

Note that for simplicity of arguments, we do not consider the total-cost-minimization algorithms given the information of unoccupied links (i.e., temporary \textit{shortest} path (TSP) algorithm and adaptive \textit{shortest} path (ASP) algorithm) whereas we use the TFP and AFP algorithms. Such algorithms yield paths that wait at a node when the cost of waiting is smaller than the additional cost of using a detour path. Therefore, the results are qualitatively expected to be interpolations of those obtained from the SSP and TFP/AFP algorithms (with the total cost larger than that of the SSP but smaller than that of the TFP/AFP, and the total time of travel shorter than that of the SSP but longer than that of the TFP/AFP). In fact, if $c_{\rm W}$ is set to a sufficiently large value, the TSP and ASP algorithms will be reduced to the TFP and AFP algorithms, respectively. Also, if $c_{\rm W}=0$ is set, they will be reduced to the SSP algorithm.

\subsection{Scenarios}\label{sec:scenario}
We simulated four different settings.
The first one was the baseline scenario, in which the system was not perturbed.
The origin and destination nodes of each packet were selected uniformly at random.
We examined how the three algorithms functioned in the three types of networks
for various numbers of packets, three capacity distributions, and two waiting cost values.

\begin{figure*}[t]
	\centering\includegraphics[width=160mm]{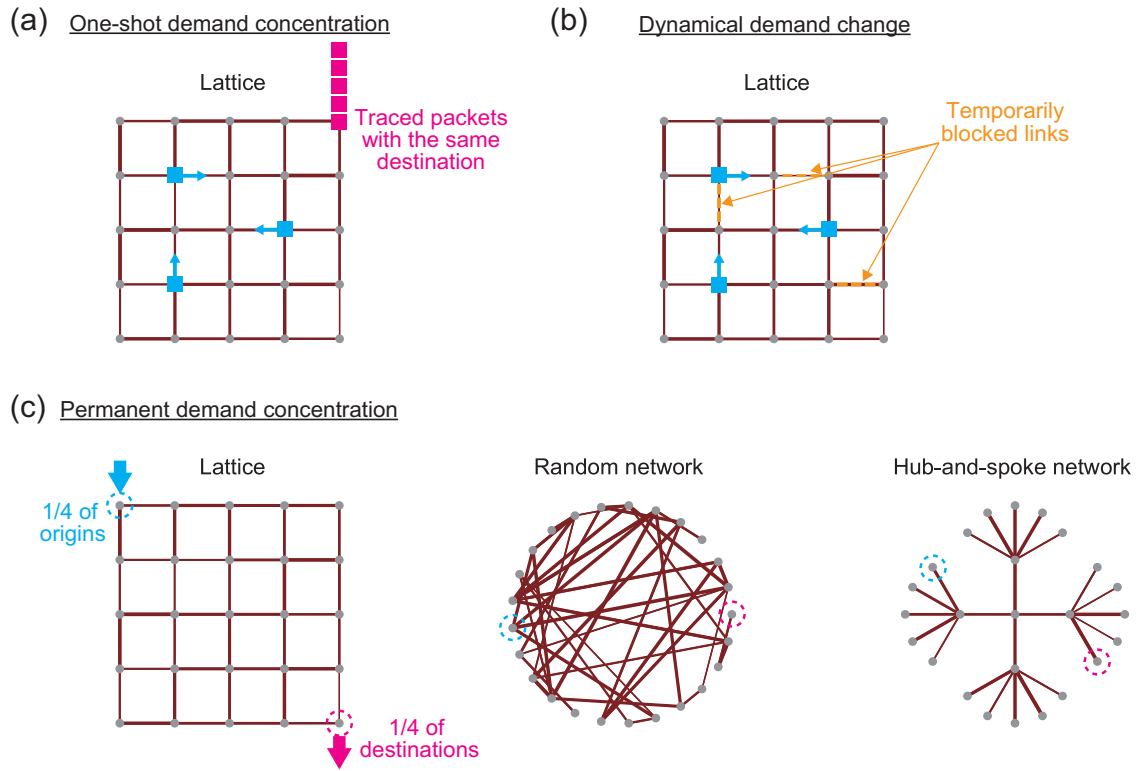}
	\caption{Scenarios with three different types of changes in demand. (a) One-shot demand concentration scenario. (b) Dynamical demand change scenario. (c) Permanent demand concentration scenario. The cyan and magenta circles represent nodes that are frequently (with a probability of 1/4) selected as origins and destinations with high demand, respectively.}
	\label{fig:concentration_sch}
\end{figure*}

The second one was the \textit{one-shot-demand-concentration} scenario, in which a set of $M$ ($=2$ or $5$ $>$ capacity) packets
that could not be sent at the same time was randomly placed in addition to the $N$ regular packets already existing in the system (Fig. \ref{fig:concentration_sch}(a)).
This elucidated the system behavior in response to a temporary and localized excess demand.
These $M$ packets had the same origins and destinations.
The origins and destinations of the packets, including the regular ones, were selected uniformly at random.
We traced these $M$ packets and measured the time taken for the first and last packets to arrive at the destination.
Each packet that completed its travel was removed from the system.
After all the $M$ packets had arrived, we generated a new set of traced packets and repeated the measurement.
The simulations were performed for the three algorithms, three types of networks, various numbers of packets,
three capacity distributions, two waiting cost values, and two packet numbers, $M$.

The third one was the \textit{dynamical-demand-change} scenario, in which a portion of links became temporally unavailable (Fig. \ref{fig:concentration_sch}(b)).
Every five time steps, we selected 12.5$\%$, 25$\%$, or 50$\%$ of the links uniformly at random
and blocked them for five time steps. We assumed that the routing algorithms could use the information
about how long the link blockage would last.
When the pattern of blockage changed, the AFP algorithm recomputed the path plans of all the packets.
A packet using the SSP algorithm had to wait until the next link becomes available if it is blocked.
The TFP algorithm took into account the blockage information available at the time (i.e., when a packet was generated) but was not flexible to new blockage patterns that will occur in the future.
The simulations were performed for the three algorithms, three types of networks, various numbers of packets,
three capacity distributions, two waiting cost values, and three fractions of link blockages.

The fourth one was the \textit{permanent-demand-concentration} scenario, in which each origin and destination of
the packets concentrated on a different single node (Fig. \ref{fig:concentration_sch}(c)).
When an origin (destination) node was to be selected, a single node was chosen with a probability of $0.25$,
while the other 24 nodes were selected with a probability of $0.75/24$ each.
This demand pattern was fixed throughout the simulation.
The simulations were performed for the three algorithms, three types of networks, various numbers of packets,
three capacity distributions, and two waiting cost values.

\subsection{Simulation conditions}
We ran these procedures for $T=1000$ time steps for each condition.
Because the cost values and one type of the network (i.e., the random network) were generated at random,
we independently generated 10 networks with cost values under the same condition and averaged the results over these 10 runs.

\clearpage
\section{Results}\label{sec:results}
\subsection{Network statistics}
\begin{figure*}[t]
	\centering\includegraphics[width=160mm]{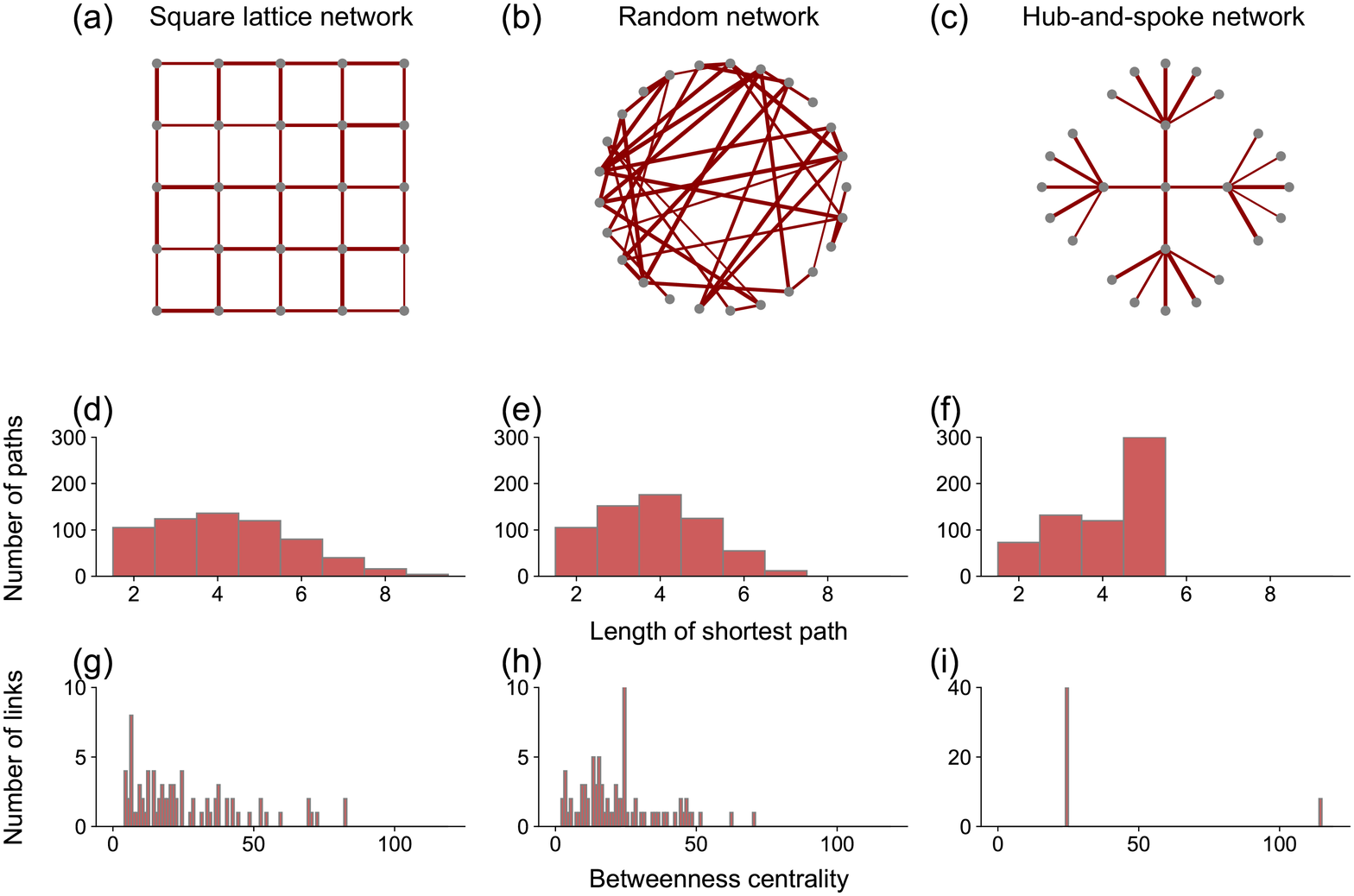}
	\caption{Network structures and statistics. (a) Square lattice network. (b) Random network. (c) Hub-and-spoke network. The thickness of the line representing each link corresponds to the cost value, $c_{ij}$ ($1\leq i \neq j \leq 25$). (d--f) Distribution of the length of the shortest path between a pair of nodes for each network. (g--i) Distribution of the link betweenness centrality for each network.}
	\label{fig:network_sch}
\end{figure*}

Here, we report the statistics of the networks (Fig.  \ref{fig:network_sch}).
Figure \ref{fig:network_sch}(d--f) shows an example of the length distribution of the shortest paths computed for all the pairs of nodes for each network. The distributions for the square lattice network (Fig. \ref{fig:network_sch}(d)) and random network (Fig. \ref{fig:network_sch}(e)) both showed peaks at length 4.
Because the square lattice network had no shortcut links, it had a larger percentage of node pairs
with longer shortest paths than did the random network.
The distribution for the hub-and-spoke network had its peak at length 5, which was the maximum value of the shortest path length (Fig. \ref{fig:network_sch}(f)).

Figure \ref{fig:network_sch}(g--i) shows the distribution of the betweenness centrality of the links,
which is a measure of how likely a link was to be used in the shortest path.
The distributions for the square lattice and random networks were similar,
but the variance was smaller in the random network (Fig. \ref{fig:network_sch}(g, h)).
Meanwhile, the betweenness centrality of the hub-and-spoke network took only two values, namely, 24 or 114,
corresponding to 40 (directional) links connecting to the peripheral nodes and 8 links connecting to the central node, respectively.
These distributions were reflected by the capacity distribution (Fig. \ref{fig:capacity_sch}(d--f)).

\subsection{Fundamental analyses of system}
\subsubsection{Routing algorithm and network structure}
We compared the total cost and travel time of packets across the three different algorithms
and waiting cost conditions for the baseline scenario for each network. Note that because the average path length, number of links, and total capacity of links differ between the three networks, we do not quantitatively compare the results between networks.

First, we report the cases with $c_{\rm W}=0$ (i.e., negligible waiting cost)
and the \textit{single-packet-per-link} capacity condition.

\begin{figure*}[t]
	\centering\includegraphics[width=160mm]{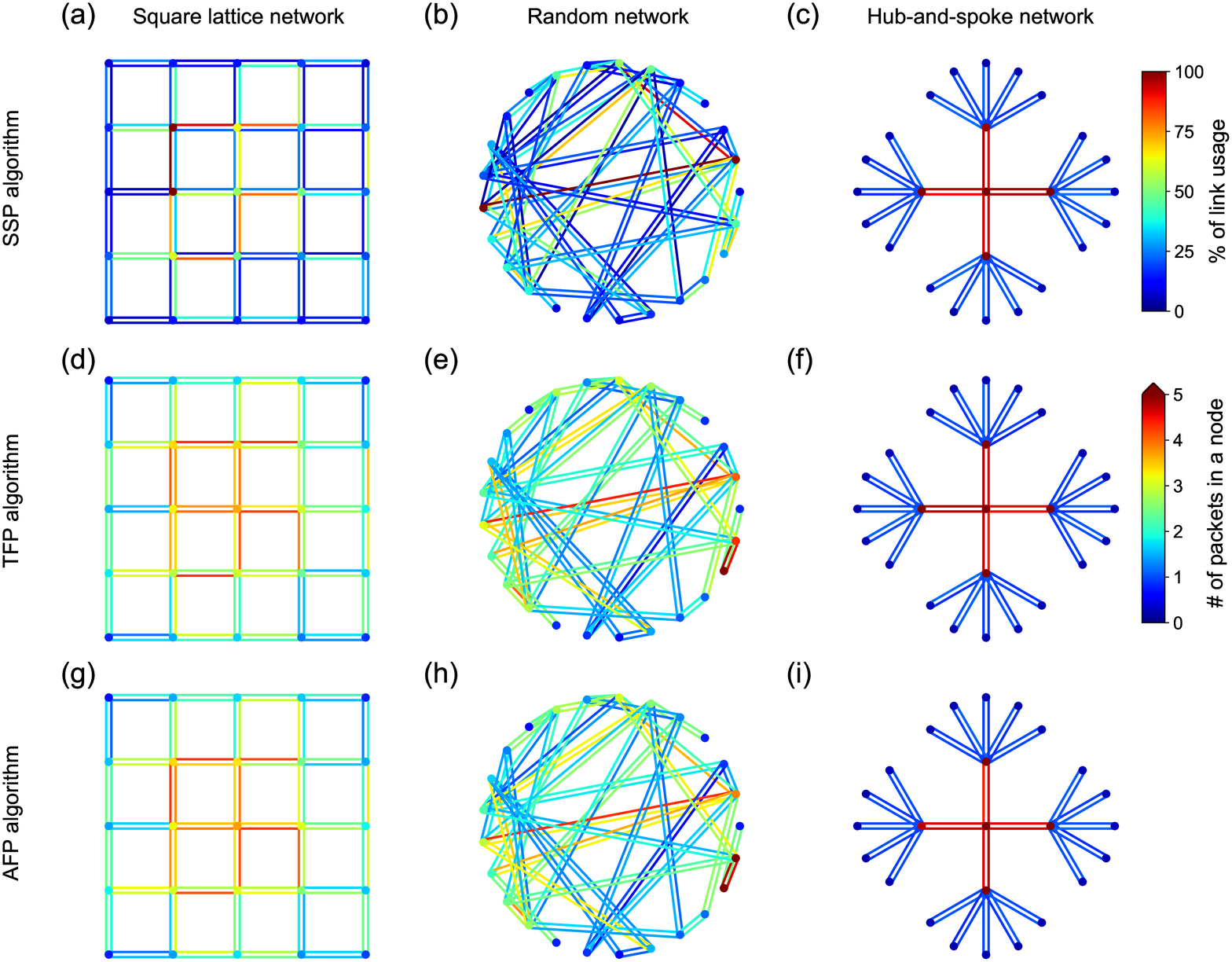}
	\caption{Examples of congestion pattern in baseline scenario with single-packet-per-link capacity distribution, $N=50$ packets, and no waiting cost (i.e., $c_{\rm W}=0$), conditioned on three algorithms and networks. (a, d, g) Square lattice network. (b, e, h) Random network. (c, f, i) Hub-and-spoke network. (a--c) SSP algorithm. (d--f) TFP algorithm. (g--i) AFP algorithm.
		The colors of the links and nodes represent the fractions of time that the links were occupied and the average numbers of packets in the nodes, respectively. The results were obtained by counting the number of packets on the links and nodes every time step and averaging them over $T=1000$ time steps of a single simulation run.}
	\label{fig:congestion_pattern}
\end{figure*}

Figure \ref{fig:congestion_pattern} illustrates how the three algorithms did (did not) even out the traffic demand in each network. In the square lattice and random networks, the TFP and AFP algorithms effectively dispersed the delivery routes. In contrast, in the hub-and-spoke network, the congestion pattern remained the same regardless of the choice of routing algorithm.

\begin{figure*}[t]
	\centering\includegraphics[width=160mm]{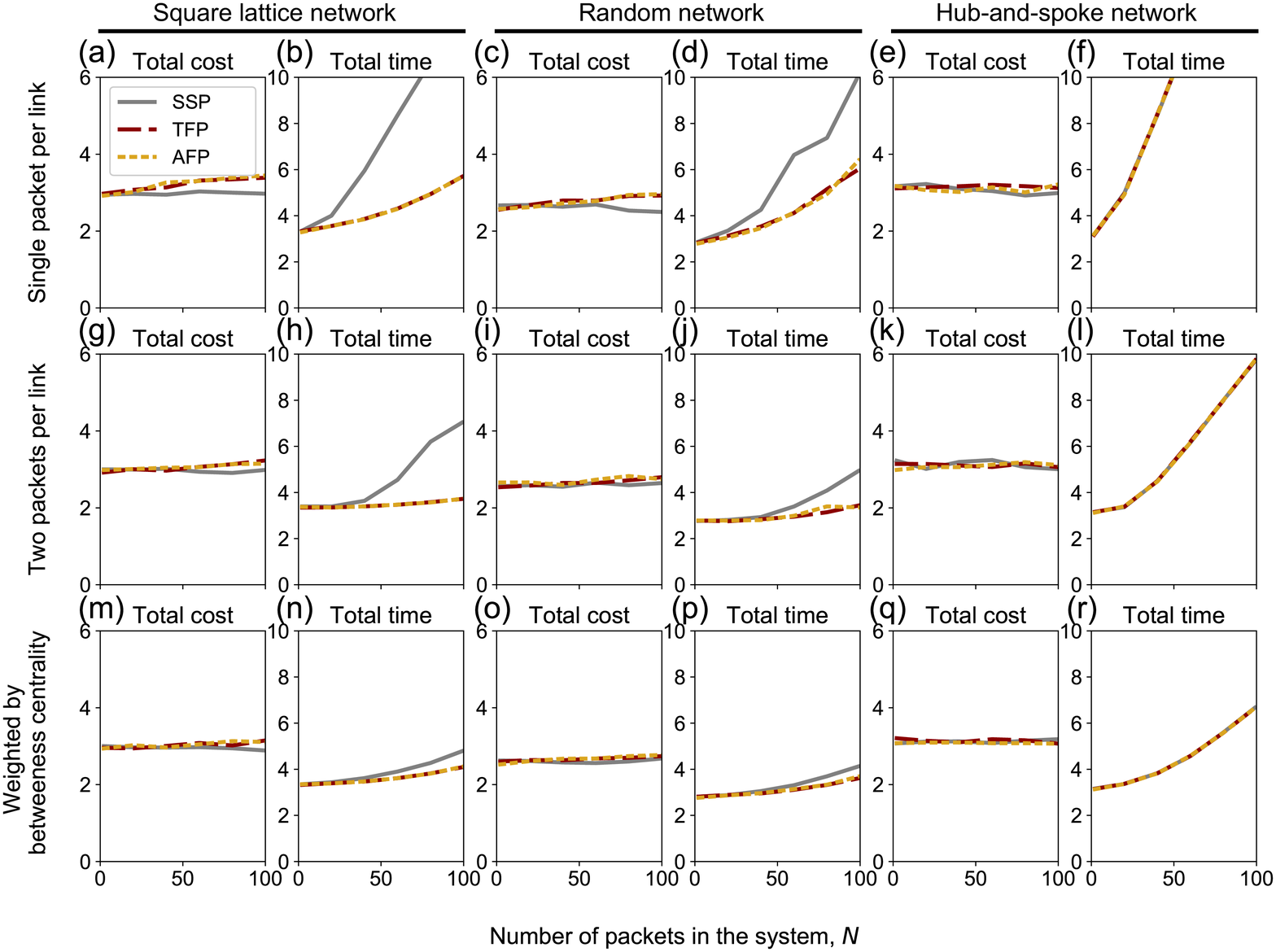}
	\caption{Total cost and time of travel in baseline scenario with no waiting cost (i.e., $c_{\rm W}=0$)
		under various conditions on algorithms, types of networks,
		capacity distributions, and number of packets in system.
		(a, b, g, h, m, n) Square lattice network.
		(c, d, i, j, o, p) Random network. (e, f, k, l, q, r) Hub-and-spoke network. (a--f) Single-packet-per-link capacity distribution.
		(g--l) Two-packets-per-link capacity distribution. (m--r) Weighted-by-betweenness-centrality capacity distribution.
		The results were averaged over $T=1000$ time steps and $10$ independent runs.}
	\label{fig:fundamental}
\end{figure*}

These properties were also reflected by the total cost and travel time of individual packets. In the square lattice and random networks, the TFP and AFP algorithms increased the travel cost slightly compared with the SSP algorithm (Fig. \ref{fig:fundamental}(a, c)),
but they significantly reduced the travel time when the number of packets in the system increased, which suggests that they were successful in finding alternative routes efficiently (Fig. \ref{fig:fundamental}(b, d)).
The TFP and AFP algorithms performed similarly.
Meanwhile, in the hub-and-spoke network, all three algorithms resulted in similar travel costs and times (Fig. \ref{fig:fundamental}(e, f)) because this network had no bypass route for any pair of nodes.
Also, for this reason, the travel cost and time rapidly increased for excess demand.

When the waiting cost was not negligible (e.g., $c_{\rm W}=0.5$), the cost of travel increased with travel time,
significantly deteriorating the performance of the SSP algorithm (Fig. \ref{fig:fundamental_cw05}(a, c)).
The presence of the waiting cost did not substantially change the travel time (Figs. \ref{fig:fundamental_cw05}(b, d, f) and \ref{fig:fundamental}(b, d, f)).
As in the cases with $c_{\rm W}=0$, the performance measures of the three algorithms were similar in the hub-and-spoke network (Fig. \ref{fig:fundamental_cw05}(e, f)).

Because the waiting cost did not change the route selecting behavior significantly,
we focus on the cases with $c_{\rm W}=0$ in the rest of this article and show the results for $c_{\rm W}=0.5$
in Appendices.

\subsubsection{Capacity distributions}
Next, we examined the effect of capacity distributions on the performance of the algorithms.
The total capacities (i.e., the sum of the capacity values over all the links in a network)
of the \textit{two-packets-per-link} and \textit{weighted-by-betweenness-centrality} conditions,
were both twice that of the \textit{single-packet-per-link} condition.

\begin{figure*}[t]
	\centering\includegraphics[width=160mm]{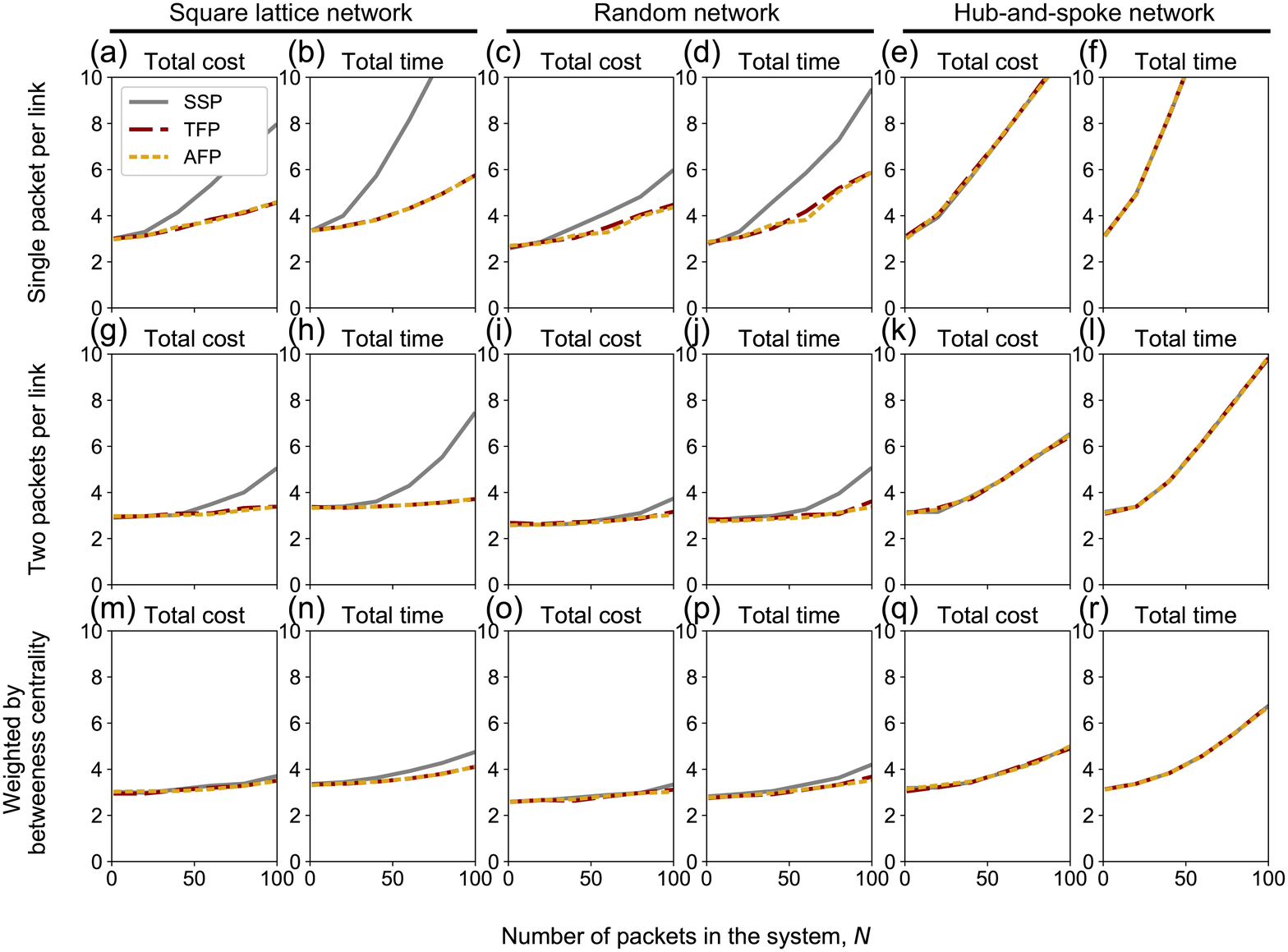}
	\caption{Total cost and time of travel in baseline scenario with positive waiting cost ($c_{\rm W}=0.5$)
		under various conditions on algorithms, types of networks,
		capacity distributions, and number of packets in system.
		(a, b, g, h, m, n) Square lattice network.
		(c, d, i, j, o, p) Random network. (e, f, k, l, q, r) Hub-and-spoke network. (a--f) Single-packet-per-link capacity distribution.
		(g--l) Two-packets-per-link capacity distribution. (m--r) Weighted-by-betweenness-centrality capacity distribution.
		The results were averaged over $T=1000$ time steps and $10$ independent runs.}
	\label{fig:fundamental_cw05}
\end{figure*}

These two capacity distributions reduced the travel time in all three networks and three algorithms
with $c_{\rm W}=0$ (Fig. \ref{fig:fundamental}) and $c_{\rm W}=0.5$ (Fig. \ref{fig:fundamental_cw05}) as compared with the \textit{single-packet-per-link} distribution.
In detail, for the square lattice random networks, they kept the travel time almost constant when the number of packets, $N$, was smaller than $\approx 50$, while it was suppressed to about 30-60$\%$ of that of the \textit{single-packet-per-link} condition when $N=100$. In the hub-and-spoke network, the rapid increase in the travel time was significantly suppressed (for example, the travel time was reduced to less than half for $N=50$).
The total cost was not changed when $c_{\rm W}=0$, but was reduced when $c_{\rm W}=0.5$ because
it was associated with the total time.

With the SSP algorithm, the total travel time decreased considerably in the square lattice and hub-and-spoke networks
under the \textit{weighted-by-betweenness-centrality} condition  (Figs. \ref{fig:fundamental}(b, h, n, f, l, r) and \ref{fig:fundamental_cw05}(b, h, n, f, l, r)), suggesting that the capacity resource was effectively allocated to important links.
In particular, the hub-and-spoke network benefited from this capacity distribution
because the demands on the links were concentrated on a small number of links (Fig. \ref{fig:network_sch}(i)).
In contrast, the difference between the two capacity conditions was not large in the random network
because it had a smaller variance of the betweenness centrality than did the other two networks; thus
the shortest paths were more dispersed (Fig. \ref{fig:network_sch}(g--i)).

The TFP and AFP algorithms performed similarly under the two capacity conditions except in the hub-and-spoke network (Figs. \ref{fig:fundamental}(g--j, m--p) and \ref{fig:fundamental_cw05}(g--j, m--p)).
These algorithms dispersed the paths without significantly increasing the cost in the square lattice and random networks, and thus, the \textit{weighted-by-betweenness-centrality} condition was not superior to the \textit{two-packets-per-link} condition.
In the hub-and-spoke network, because these two algorithms generated paths similar to those of the SSP algorithm,
the \textit{weighted-by-betweenness-centrality} capacity distribution resulted in a larger decrease in the travel time
(Figs. \ref{fig:fundamental}(f, l, r) and \ref{fig:fundamental_cw05}(f, l, r)).

\subsection{One-shot demand concentration}
\begin{figure*}[t]
	\centering\includegraphics[width=160mm]{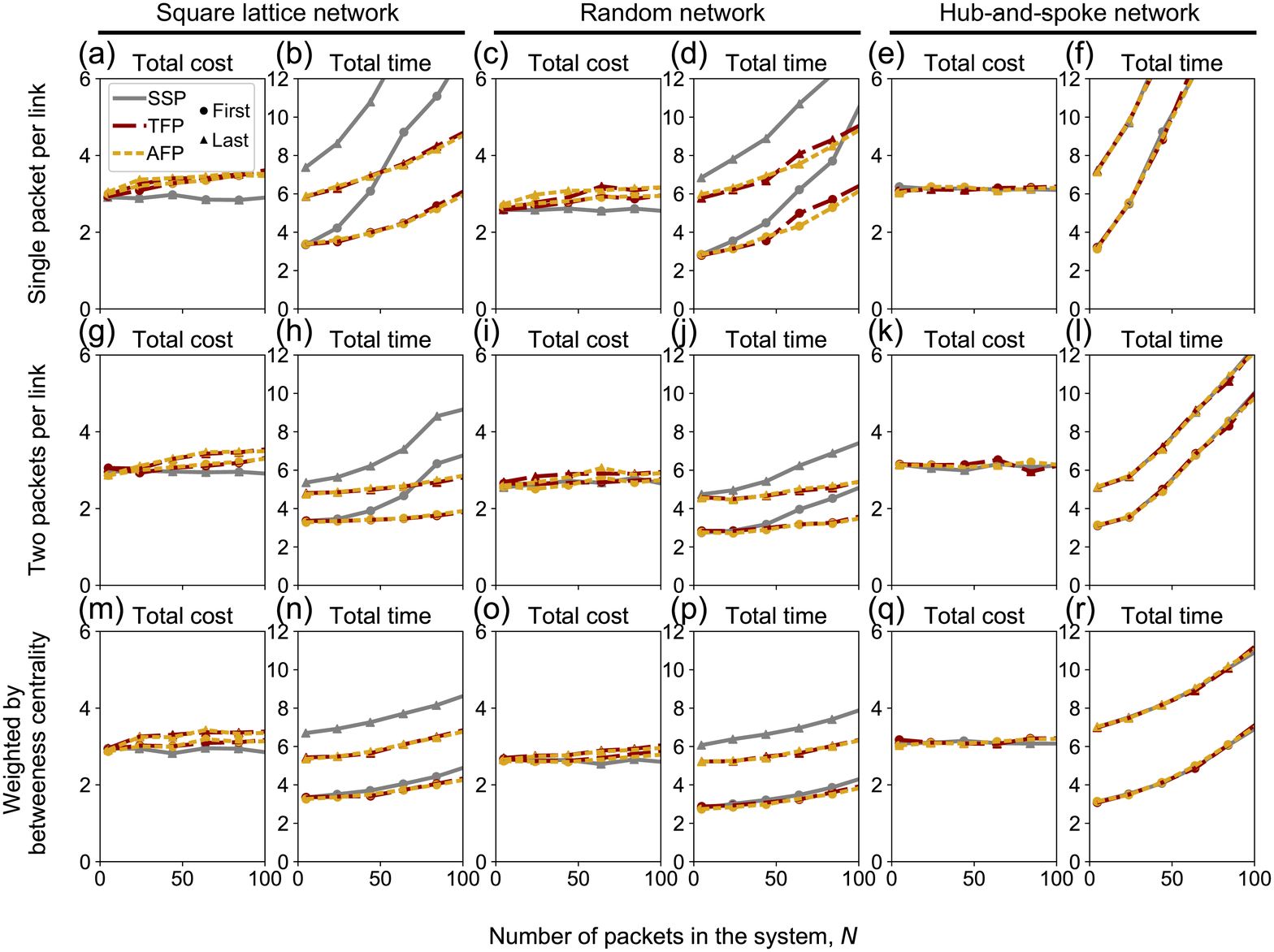}
	\caption{Total cost and time of travel in one-shot demand concentration scenario with no waiting cost ($c_{\rm W}=0$)
		under various conditions on algorithms, types of networks,
		capacity distributions, and number of packets in system.
		(a, b, g, h, m, n) Square lattice network.
		(c, d, i, j, o, p) Random network. (e, f, k, l, q, r) Hub-and-spoke network. (a--f) Single-packet-per-link capacity distribution.
		(g--l) Two-packets-per-link capacity distribution. (m--r) Weighted-by-betweenness-centrality capacity distribution.
		The results were averaged over $T=1000$ time steps and $10$ independent runs.}
	\label{fig:oneshot}
\end{figure*}

Next, we examine the response of the system to a temporary and localized increase in demand (i.e., the \textit{one-shot-demand-concentration} scenario).
We placed a set of $M$ (= 2 or 5) packets having the same destination at a random position  (Fig. \ref{fig:concentration_sch}(a)) and recorded
the travel paths of the first and last packets to arrive.
Figure \ref{fig:oneshot} shows the total costs and travel times of the
first and last packets for $c_{\rm W} =0$ and $M=5$.

Because the SSP algorithm generated the same paths for the set of traced packets,
the total cost was the same between the first and last packets in all the networks.
However, as the last packet had to wait until the previous packets had cleared out and the shortest path became available,
the difference in the arrival times between the first and last packets was large (Fig. \ref{fig:oneshot}).
The difference was the smallest in the \textit{two-packets-per-link} capacity distribution, while the two other capacity
distributions yielded similar time differences.
This is in stark contrast to the baseline scenario, in which the \textit{weighted-by-betweenness-centrality} capacity distribution
performed the best in the majority of cases (Figs. \ref{fig:oneshot} and \ref{fig:oneshot_cw05}).
The time difference was not substantially influenced by the number of packets in the system in the range of $1\leq N \leq 100$.

The TFP and AFP algorithms reduced the difference in the arrival time in the square lattice and
random networks (Fig. \ref{fig:oneshot}(b, d, h, j, n, p)) by dispersing the paths, thereby increasing
the effective capacity of delivery from the origin to the destination. The difference in arrival time was the smallest in the
\textit{two-packets-per-link} capacity distribution because detour routes were likely to exist everywhere,
as the capacity resources were not concentrated on a limited number of links.
The total travel cost was only slightly greater for the last packet than for the first (Fig. \ref{fig:oneshot}(a, c, g, i, m, o)).
In the hub-and-spoke network, the two algorithms yielded results similar to those of the SSP algorithm because no detour routes existed.

The pattern of the results remained the same when we set $c_{\rm W}=0.5$ (Fig. \ref{fig:oneshot_cw05})
except that the total cost increased with the travel time.
When we set $M=2$ (and $c_{\rm W}=0$), the results were qualitatively unchanged, but
the difference in the travel time between the first and last (i.e., second) packets was small (Fig. \ref{fig:oneshot_packet_size2}).

\subsection{Dynamical demand change}
We considered a situation where some links were randomly closed for a certain length of time for external reasons (i.e., \textit{dynamical-demand-change} scenario; Fig. \ref{fig:concentration_sch}(b); see Sec. \ref{sec:scenario} for details).
Link closures that may occur in the future were not taken into account by the routing algorithms
(but information about current link closures, i.e., which links were being closed and until when they were closed, was available for the TFP and AFP algorithms).
We simulated the system dynamics with three percentages of blocked links, i.e., 12.5$\%$, 25$\%$, and 50$\%$.

\begin{figure*}[t]
	\centering\includegraphics[width=160mm]{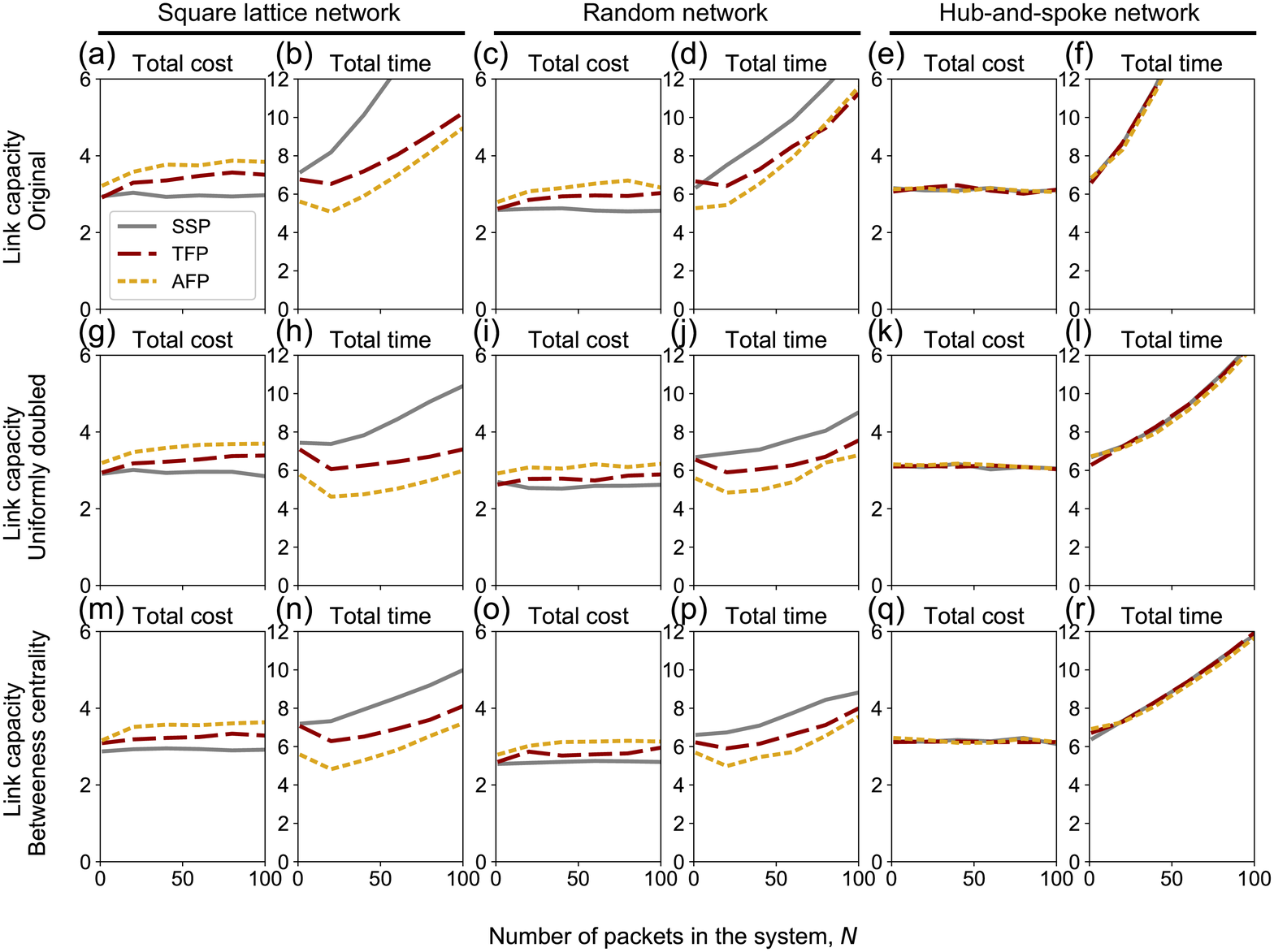}
	\caption{Total cost and time of travel in dynamical demand change scenario
		with 25$\%$ of links being closed and no waiting cost ($c_{\rm W}=0$)
		under various conditions on algorithms, types of networks,
		capacity distributions, and number of packets in system.
		(a, b, g, h, m, n) Square lattice network.
		(c, d, i, j, o, p) Random network. (e, f, k, l, q, r) Hub-and-spoke network. (a--f) Single-packet-per-link capacity distribution.
		(g--l) Two-packets-per-link capacity distribution. (m--r) Weighted-by-betweenness-centrality capacity distribution.
		The results were averaged over $T=1000$ time steps and $10$ independent runs.}
	\label{fig:dynamical}
\end{figure*}

Figure \ref{fig:dynamical} shows the total cost and travel time when 25 $\%$ of the links were closed and $c_{\rm W}=0$.
With the SSP algorithm, the travel time increased compared with that in the baseline scenario (Fig. \ref{fig:fundamental}).
The results were similar between the \textit{two-packets-per-link} and \textit{weighted-by-betweenness-centrality} capacity conditions because of the large negative impact of the closure of a critical large-capacity link, which offset the advantage of
the \textit{weighted-by-betweenness-centrality} capacity distribution.

The TFP and AFP algorithms both reduced the travel times in the square lattice and random networks (Fig. \ref{fig:dynamical}).
In particular, the AFP algorithm yielded faster travels than did the TFP algorithm
by adaptively rerouting the packets.
The performance measures of these algorithms were similar between the \textit{two-packets-per-link} and \textit{weighted-by-betweenness-centrality} capacity conditions.
The total cost was larger for the AFP, TFP, and SSP algorithms, in that order.
In the hub-and-spoke network, no substantial difference was found between these algorithms.

The results remained qualitatively the same for the different percentages of link closure (i.e., $12.5\%$ and $50\%$; Figs. \ref{fig:dynamical_12_5} and \ref{fig:dynamical_50}, respectively) and a different value of waiting cost (i.e., $c_{\rm W}=0.5$; Fig. \ref{fig:dynamical_25_cw05}).

\subsection{Permanent demand concentration}
\begin{figure*}[t]
	\centering\includegraphics[width=160mm]{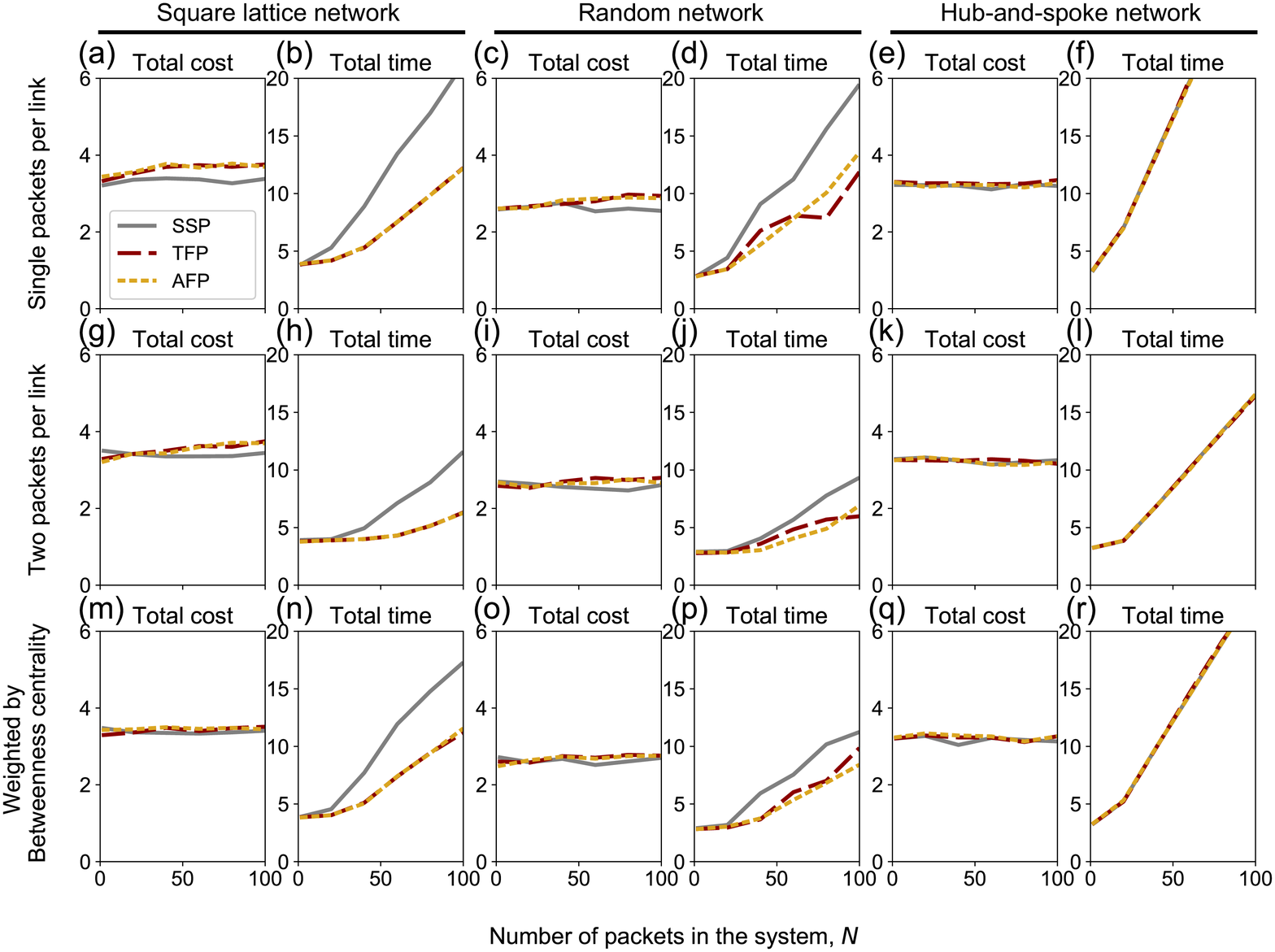}
	\caption{Total cost and time of travel in permanent demand change scenario with no waiting cost ($c_{\rm W}=0$)
		under various conditions on algorithms, types of networks,
		capacity distributions, and number of packets in system.
		(a, b, g, h, m, n) Square lattice network.
		(c, d, i, j, o, p) Random network. (e, f, k, l, q, r) Hub-and-spoke network. (a--f) Single-packet-per-link capacity distribution.
		(g--l) Two-packets-per-link capacity distribution. (m--r) Weighted-by-betweenness-centrality capacity distribution.
		The results were averaged over $T=1000$ time steps and $10$ independent runs.}
	\label{fig:permanent}
\end{figure*}

Finally, we examined the permanent change in the demand pattern (i.e., the \textit{permanent-demand-concentration} scenario; Fig. \ref{fig:concentration_sch}(c); Sec. \ref{sec:scenario}).
The permanent demand concentration caused significant increases in the cost and travel time in all the cases (Figs. \ref{fig:permanent} and \ref{fig:permanent_cw05}).
The TFP and AFP algorithms effectively dispersed the travel paths, thus reducing the travel time
compared with the SSP algorithm in the square lattice and random networks.
In the hub-and-spoke network, the three algorithms yielded similar results.
The TFP and AFP algorithms performed similarly in all the networks.

The travel time was the smallest in the \textit{two-packets-per-link} capacity distribution (Figs. \ref{fig:permanent}(g--l) and \ref{fig:permanent_cw05}(g--l)),
because the uniform capacity distribution allowed detour routes to be found even when the demand pattern was changed.

\section{Conclusion}\label{sec:conclusion}
We examined the effect of network topology on the performance of the network logistics system under perturbations.
The following results were obtained consistently for the three demand change scenarios: (i) Adaptive algorithms (i.e., TFP and AFP algorithms) performed better than did the SSP algorithm except in the hub-and-spoke network, where no difference was found between all the algorithms. (ii) The capacity distribution based on the betweenness centrality (i.e., \textit{weighted-by-betweenness-centrality}) was more effective than the uniform capacity distribution with the same total capacity (i.e., \textit{two-packets-per-link}) in reducing the travel time when the demand was generated uniformly at random. In contrast, under the demand change scenarios, the \textit{two-packets-per-link} distribution reduced the travel time more than did the \textit{weighted-by-betweenness-centrality} condition.
(iii) The performance of the square lattice and random networks were qualitatively similar. These results collectively suggest that networks with redundancy can respond well to changes in demand, while hub-and-spoke networks (which do not have such redundancy) cannot benefit of the adaptive delivery operations.

For redundant networks (i.e., square lattice and random networks), we also found the following results specific to the scenarios:
(i) In the one-shot-demand-concentration scenario, the TFP and AFP algorithms effectively reduced the difference between the arrival times of the first and last packets of a set of packets having the same origins, destinations, and departure times. This quantity would be a useful measure for evaluating the performance of logistics network.
(ii) In the dynamical-demand-change scenario, the AFP algorithm effectively found a temporally better route than the TFP algorithm, which did not change the route during the travel.

In this paper, for simplicity of discussion, we focused on the total cost and time of travel as measures of performance.
However, other measures, including sustainability ones \citep{Montreuil2011-hh}, should be considered in more detailed system evaluation. For example, studies have been shown that the PI can reduce green gas emissions from 14\% to 50\% \citep{Pan2013-xd}.
In our model, the length of travel is roughly measured by the total cost (when we set $c_{\rm w}=0$). We found that travel time can be reduced in redundant networks without a substantial increase in the total cost, suggesting the efficiency of the algorithms in terms of sustainability.
In addition, we have only partially considered the cost of inventory by using a waiting cost. By describing it in more detail, we can also consider the problem of coordinated inventory management \citep{Darmawan2021-pl} in logistics networks with our approach.

It should be noted that we set the cost of using a link at random in the range of 0.5 to 1.5. In practice, the cost of travel may significantly differ between links if multimodal transportations and/or different types of players (e.g., manufacturers, wholesalers and retailers) are to be included. The same issue applies to the cost of waiting, $c_{\rm W}$. In this paper, we did not pursue realistic settings for these values, focusing instead on the general behavior of the system independent of the specific parameter settings. We believe that the main conclusion of the paper (i.e., redundant networks are robust) is not substantially influenced by the choice of the parameter values unless one sets extreme values.
Also, in our model, the number of packets sent in a single transport on a link was limited to a small non-negative integer,
while many packets can be sent at once in real logistics systems. We believe that this does not substantially affect our conclusions.
Even if the link capacities and number of packets in the system were increased, a network with bypass routes would have still allowed for the efficient dispersal and delivery of packets in case of excess demand.

In order to provide baseline results as a first step, the efficiency gains in logistics due to consolidation (and other delivery constraints) were not considered in the current model, but their impacts should be carefully examined in the future. For example, if an operation that increases the load factor of a single delivery by consolidating the transportation of one link to another link is possible, the routing algorithm will be modified. However, even in such a case, the shortest or fastest route would be the baseline choice, and the overall pattern of logistics is not expected to change significantly. From the standpoint of load factor, hub-and-spoke networks enable delivery with a high load factor by accumulating the freight from subordinate distribution centers. On the other hand, networks with many bypass routes cannot benefit from accumulation, but as the number of delivery route patterns is large, there would be many chances of consolidation between different delivery routes. Comparing these two types of networks under various conditions will be an interesting research topic in the future.

In this paper, we examined three types of capacity distributions but did not try to determine the best one.
Because the packets interacted with each other's routes, selection of the best capacity distribution was
not a straightforward process.
For example, a previous study indicated that traffic congestions cannot be predicted simply from the betweenness centrality \citep{Holme2003-ig}. The results will also vary depending on which measure of robustness \citep{Gu2020-bd} is adopted.
Therefore, identification of an efficient capacity allocation given the network structure and operation of a logistics system is a subject of future research.

Our results suggest that hub-and-spoke networks are not robust to demand changes and cannot benefit from the flexible routing in a logistic network in general.
However, such a structure is beneficial in terms of (static) efficiency and is quite widely used in practice \citep{Abdinnour-Helm1999-ke}.
Thus, ways of modifying a part of the network to increase its robustness efficiently is a highly interesting subject for future research. From managerial perspective, our findings suggest that establishing regular services between distribution centers and/or building new distribution centers which allow bypass routes in the network are preferable.

The robustness of a supply chain network is closely related to the concept of resilience
\citep{Ponomarov2009-fr, Pettit2010-mm, Tukamuhabwa2015-ff, Ivanov2017-hl, Hosseini2019-mv, Dolgui2020-yf,Chen2021-ez, Ulusan2021-mw}.
Resilience is the capacity of an enterprise that enables it ``\textit{to survive, adapt, and grow in the face of turbulent change}'' \citep{Fiksel2006-oa}.
For logistics networks to be resilient, its long-term stability should be guaranteed, and this
is an important issue that needs to be clarified in the future.

Another direction for future research is multimodal transportation networks \citep{Agamez-Arias2017-iy, Crainic2018-ys, Zhou2021-ws}. In order to understand the multi-modal logistics network, it will be necessary to understand the entire network, which is composed of networks corresponding to each mode with different characteristics. We believe that our approach will provide useful insights into such issues.

\nolinenumbers

\clearpage

\clearpage

\clearpage

\clearpage

\clearpage

\clearpage

\clearpage

\clearpage

\clearpage

\clearpage

\clearpage

\section*{Appendix}
\renewcommand{\thefigure}{A\arabic{figure}}
\setcounter{figure}{0}

\clearpage
\begin{figure*}[t]
	\centering\includegraphics[width=160mm]{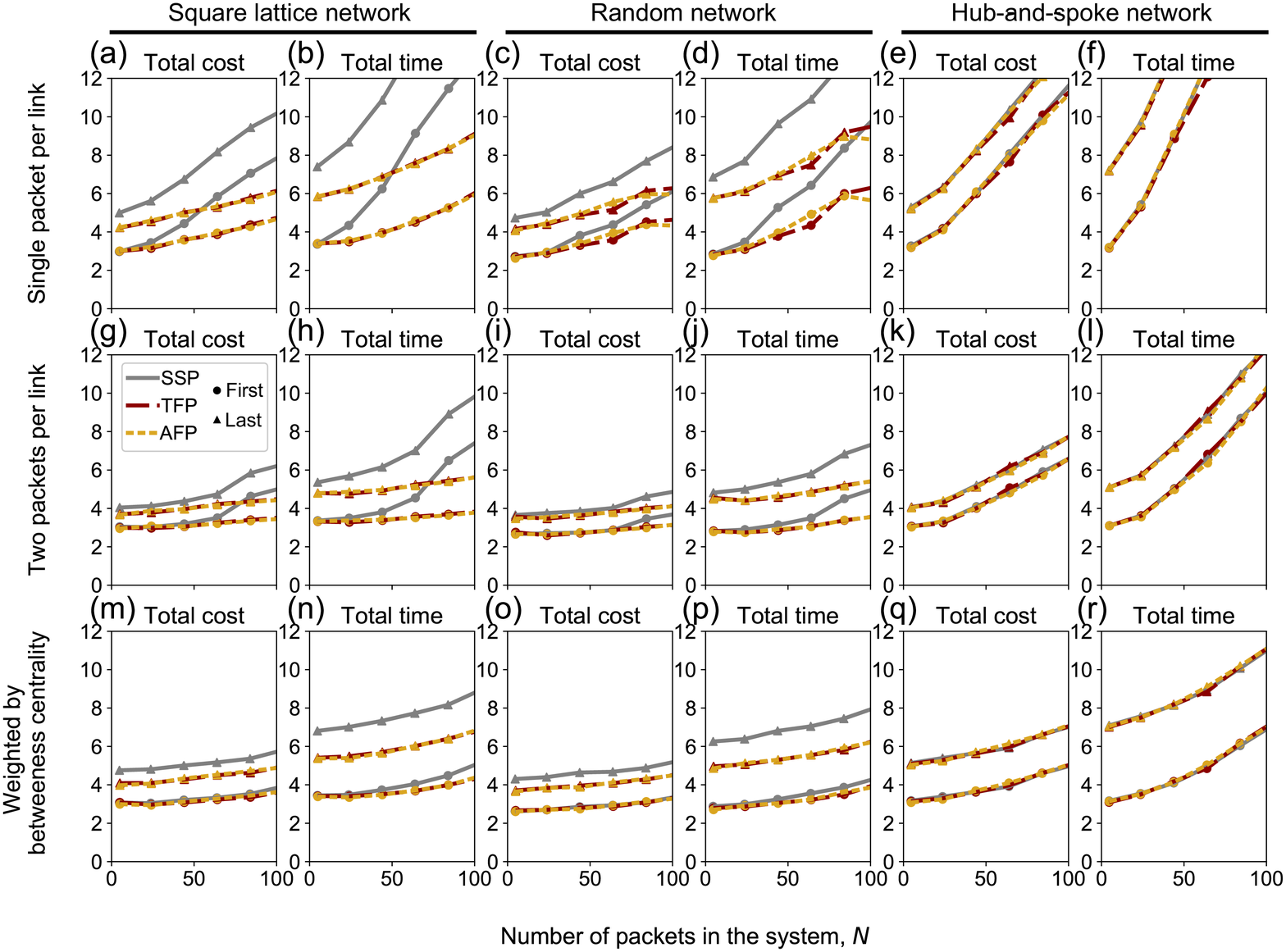}
	\caption{Total cost and time of travel in the one-shot demand concentration scenario with positive waiting cost ($c_{\rm W}=0.5$)
		under various conditions on algorithms, types of networks,
		capacity distributions, and number of packets in system.
		(a, b, g, h, m, n) Square lattice network.
		(c, d, i, j, o, p) Random network. (e, f, k, l, q, r) Hub-and-spoke network. (a--f) Single-packet-per-link capacity distribution.
		(g--l) Two-packets-per-link capacity distribution. (m--r) Weighted-by-betweenness-centrality capacity distribution.
		The results were averaged over $T=1000$ time steps and $10$ independent runs.}
	\label{fig:oneshot_cw05}
\end{figure*}

\clearpage
\begin{figure*}[t]
	\centering\includegraphics[width=160mm]{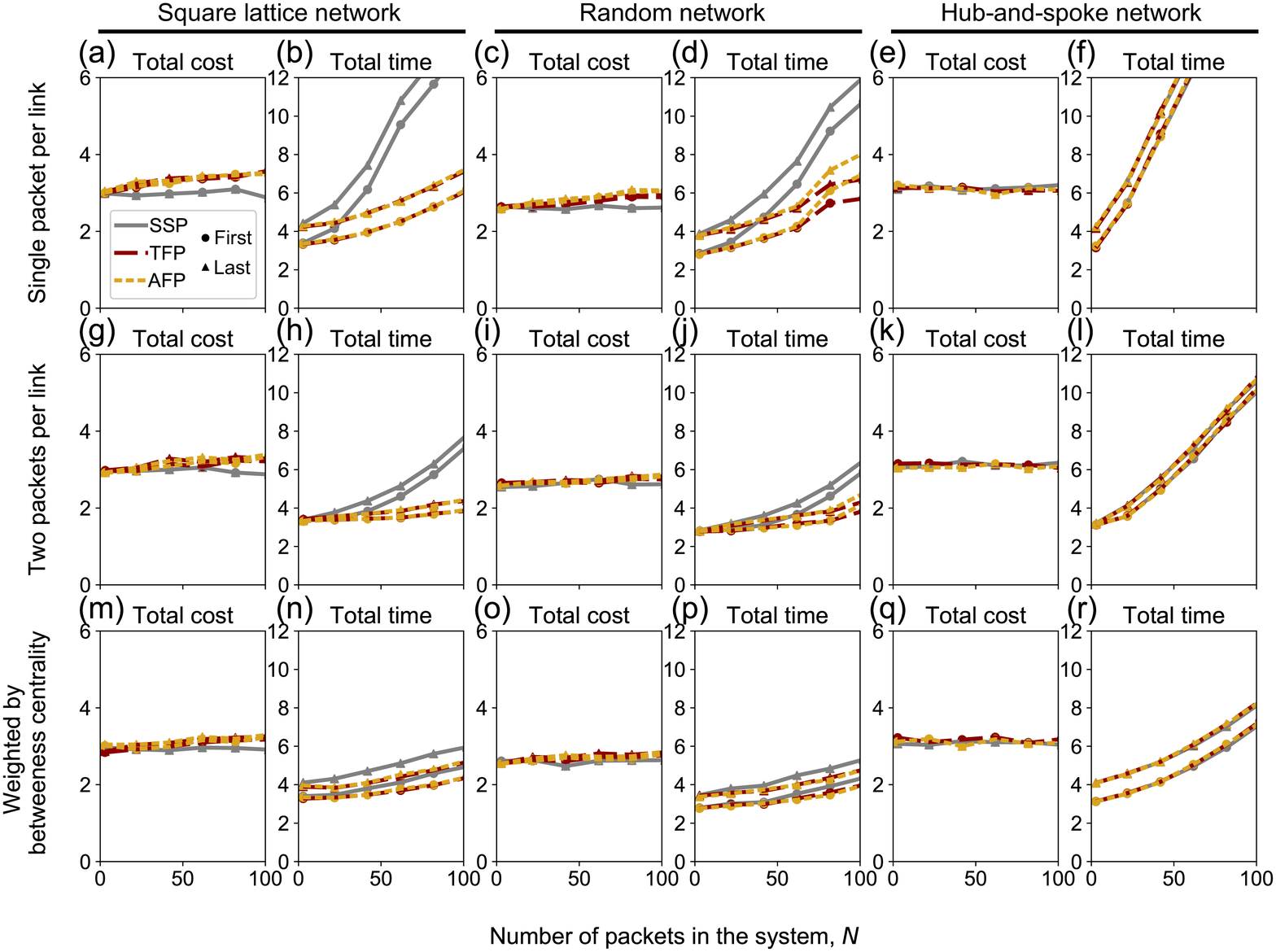}
	\caption{Total cost and time of travel in the one-shot demand concentration scenario with $M=2$
		and no waiting cost ($c_{\rm W}=0$)
		under various conditions on algorithms, types of networks,
		capacity distributions, and number of packets in system.
		(a, b, g, h, m, n) Square lattice network.
		(c, d, i, j, o, p) Random network. (e, f, k, l, q, r) Hub-and-spoke network. (a--f) Single-packet-per-link capacity distribution.
		(g--l) Two-packets-per-link capacity distribution. (m--r) Weighted-by-betweenness-centrality capacity distribution.
		The results were averaged over $T=1000$ time steps and $10$ independent runs.}
	\label{fig:oneshot_packet_size2}
\end{figure*}

\clearpage
\begin{figure*}[t]
	\centering\includegraphics[width=160mm]{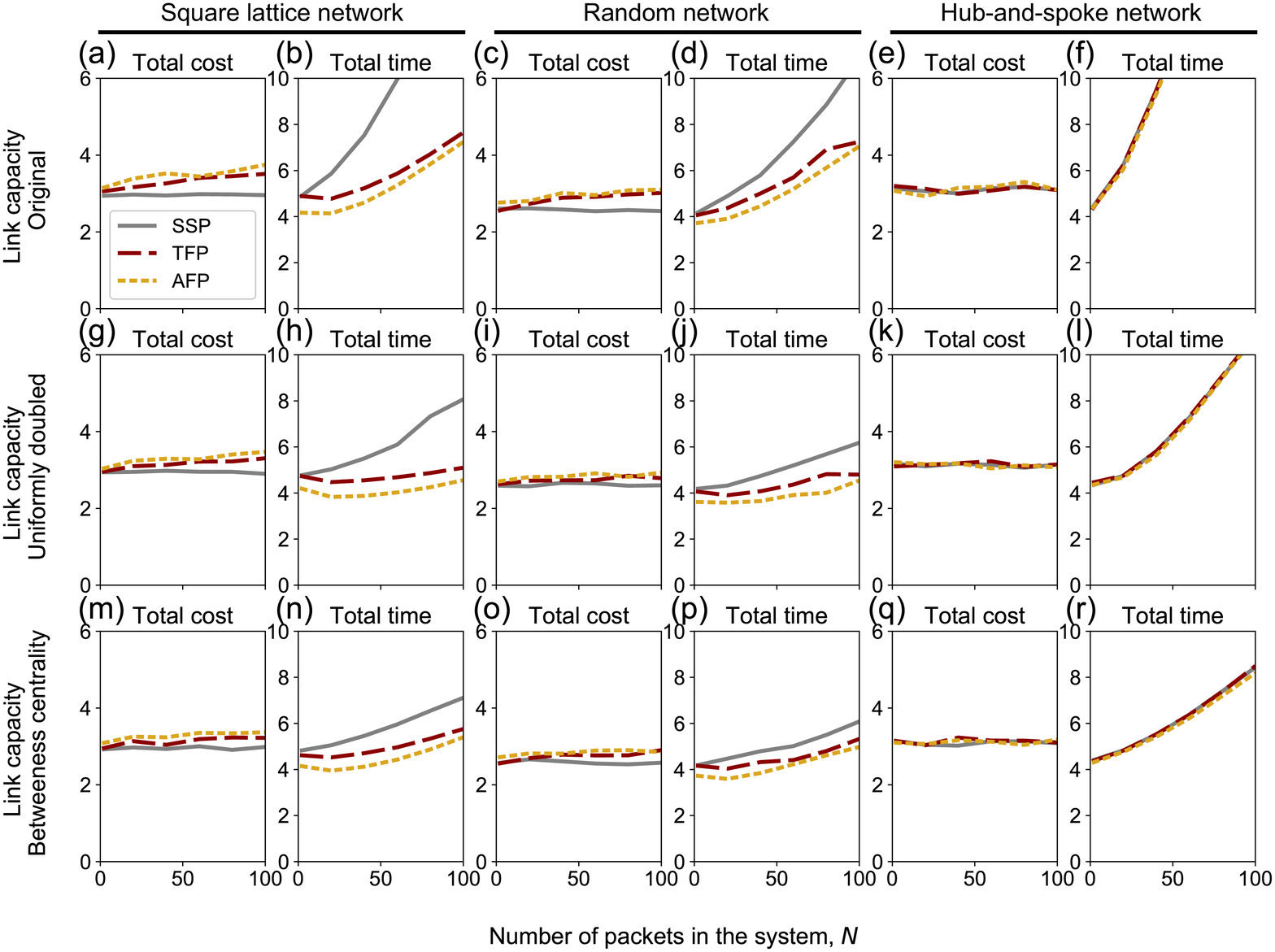}
	\caption{Total cost and time of travel in the dynamical demand change scenario with $12.5\%$ of links being closed
		and no waiting cost ($c_{\rm W}=0$) under various conditions on algorithms, types of networks,
		capacity distributions, and number of packets in system.
		(a, b, g, h, m, n) Square lattice network.
		(c, d, i, j, o, p) Random network. (e, f, k, l, q, r) Hub-and-spoke network. (a--f) Single-packet-per-link capacity distribution.
		(g--l) Two-packets-per-link capacity distribution. (m--r) Weighted-by-betweenness-centrality capacity distribution.
		The results were averaged over $T=1000$ time steps and $10$ independent runs.}
	\label{fig:dynamical_12_5}
\end{figure*}

\clearpage
\begin{figure*}[t]
	\centering\includegraphics[width=160mm]{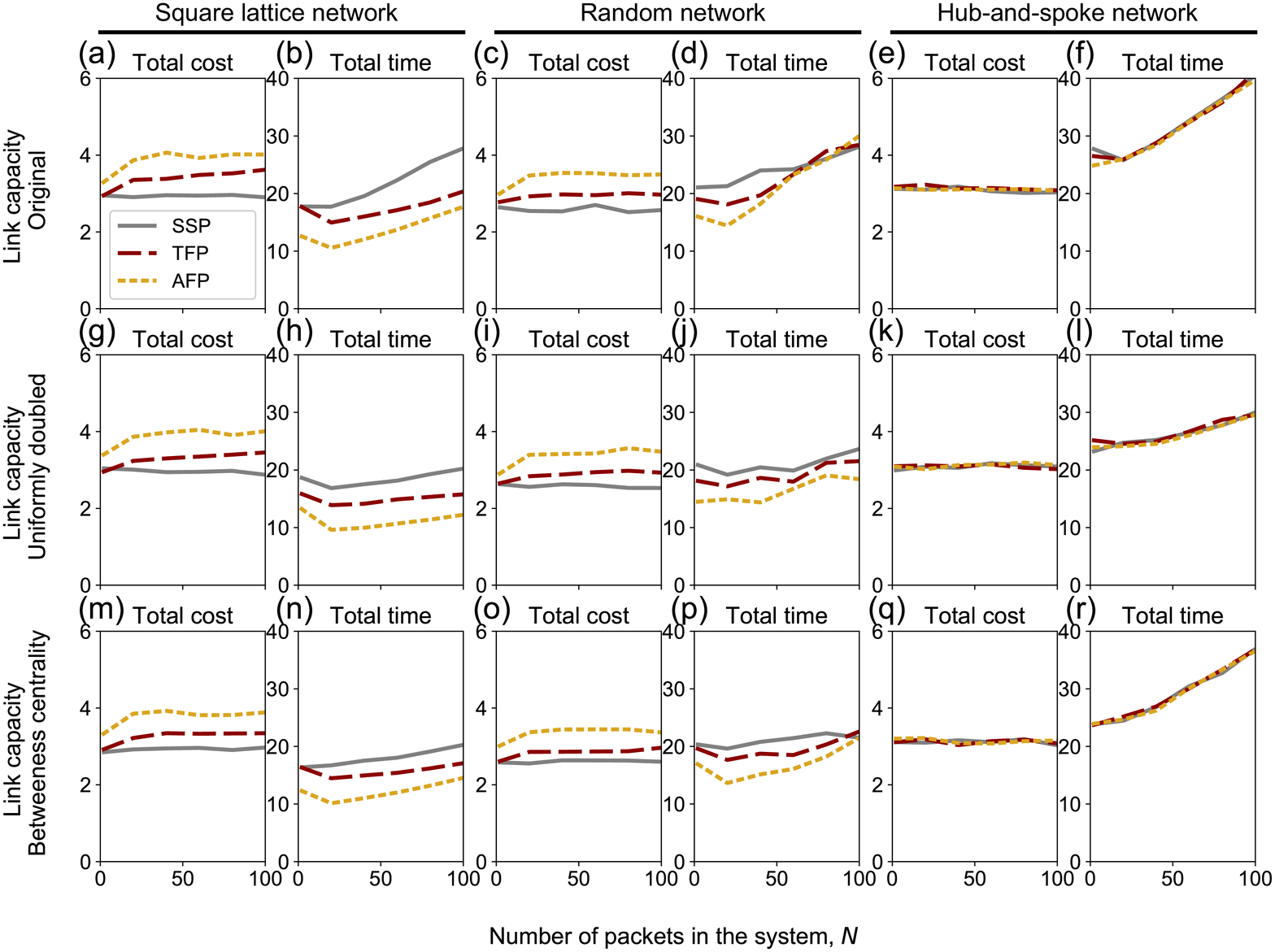}
	\caption{Total cost and time of travel in the dynamical demand change scenario with $50\%$ of links being closed
		and no waiting cost ($c_{\rm W}=0$) under various conditions on algorithms, types of networks,
		capacity distributions, and number of packets in system.
		(a, b, g, h, m, n) Square lattice network.
		(c, d, i, j, o, p) Random network. (e, f, k, l, q, r) Hub-and-spoke network. (a--f) Single-packet-per-link capacity distribution.
		(g--l) Two-packets-per-link capacity distribution. (m--r) Weighted-by-betweenness-centrality capacity distribution.
		The results were averaged over $T=1000$ time steps and $10$ independent runs.}
	\label{fig:dynamical_50}
\end{figure*}

\clearpage
\begin{figure*}[t]
	\centering\includegraphics[width=160mm]{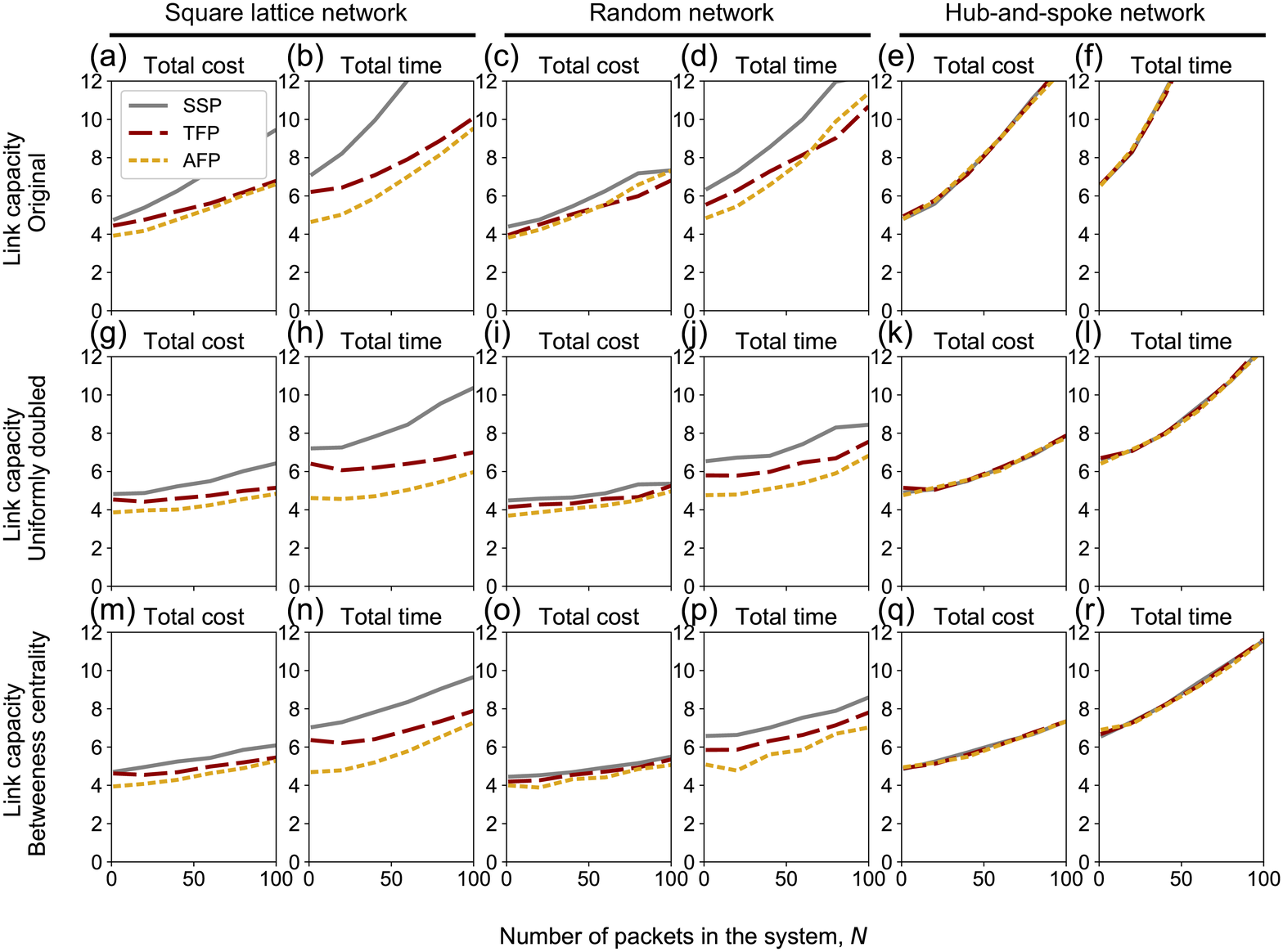}
	\caption{Total cost and time of travel in the dynamical demand change scenario with $25\%$ of links being closed
		and positive waiting cost ($c_{\rm W}=0.5$) under various conditions on algorithms, types of networks,
		capacity distributions, and number of packets in system.
		(a, b, g, h, m, n) Square lattice network.
		(c, d, i, j, o, p) Random network. (e, f, k, l, q, r) Hub-and-spoke network. (a--f) Single-packet-per-link capacity distribution.
		(g--l) Two-packets-per-link capacity distribution. (m--r) Weighted-by-betweenness-centrality capacity distribution.
		The results were averaged over $T=1000$ time steps and $10$ independent runs.}
	\label{fig:dynamical_25_cw05}
\end{figure*}

\clearpage
\begin{figure*}[t]
	\centering\includegraphics[width=160mm]{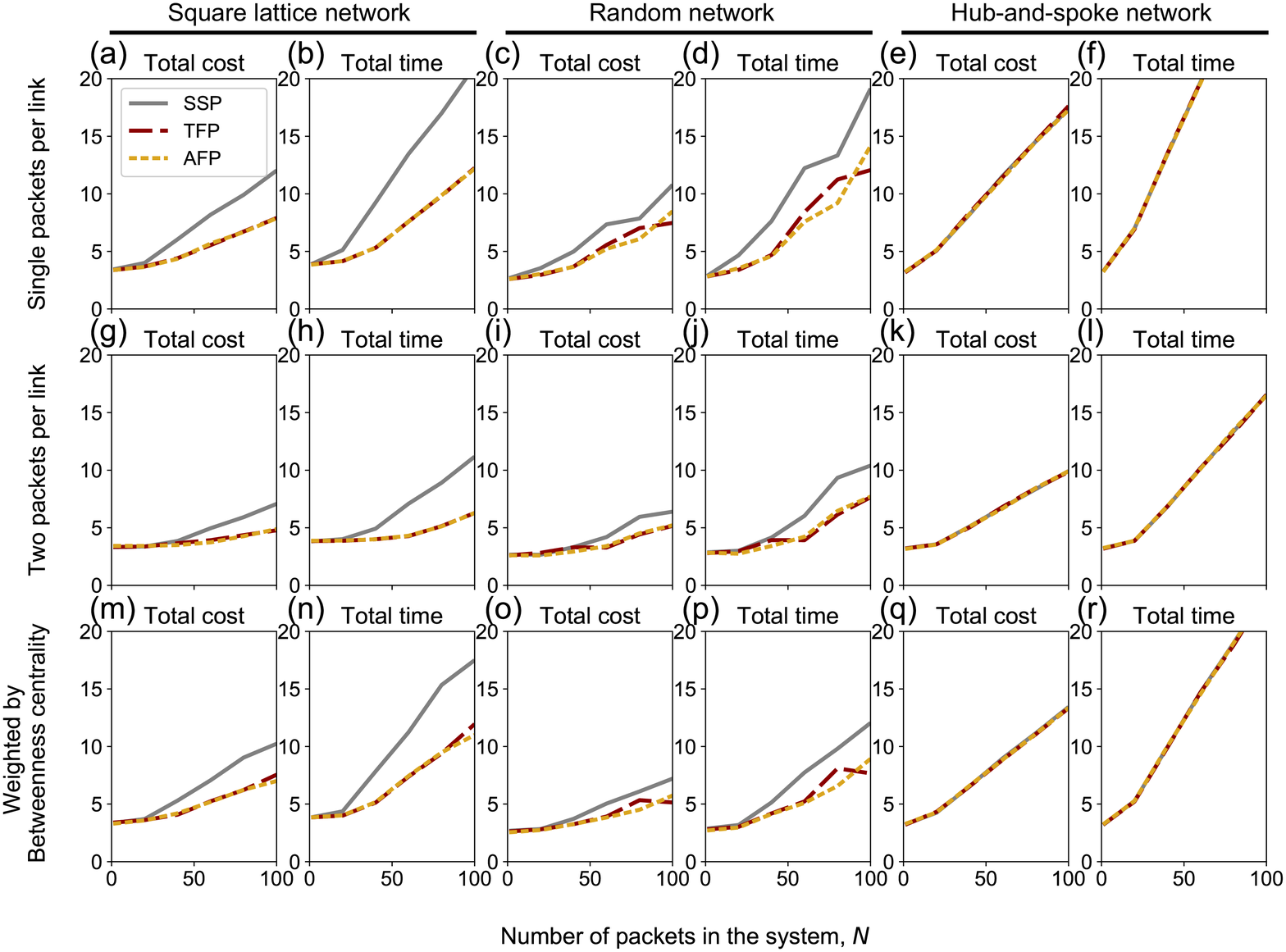}
	\caption{Total cost and time of travel in the permanent demand change scenario with positive waiting cost ($c_{\rm W}=0.5$) under various conditions on algorithms, types of networks,
		capacity distributions, and number of packets in system.
		(a, b, g, h, m, n) Square lattice network.
		(c, d, i, j, o, p) Random network. (e, f, k, l, q, r) Hub-and-spoke network. (a--f) Single-packet-per-link capacity distribution.
		(g--l) Two-packets-per-link capacity distribution. (m--r) Weighted-by-betweenness-centrality capacity distribution.
		The results were averaged over $T=1000$ time steps and $10$ independent runs.}
	\label{fig:permanent_cw05}
\end{figure*}

\end{document}